# An application of the HeunB function

S.G.Kamath*,+,#

Department of Mathematics, Indian Institute of Technology Madras

Chennai 600 036, India

*Retired, +e-mail: gkamath01865@gmail.com, #[0000-0002-1148-3787]

Abstract: How does the inclusion of the gravitational potential alter an otherwise exact quantum mechanical result? This question motivates this report, with the answer determined from an edited version of problem #12 on p.273 of Ref.1. To elaborate, we begin with the Hamiltonian associated with the system of two masses in the problem obeying Hooke's law and vibrating about their equilibrium positions in one dimension; the Schrodinger equation for the reduced mass is then solved to obtain the parabolic cylinder functions $D_\mu(x)$ as eigen functions, and the eigenvalues of the reduced Hamiltonian are calculated exactly. Parenthetically, the quantum mechanics of a bounded linear harmonic oscillator was perhaps first studied by Auluck and Kothari[2]. The introduction of the gravitational potential in the aforesaid Schrodinger equation alters the eigenfunctions $D_\mu(x)$ to the biconfluent $HeunB(q, \alpha, \gamma, \delta, \varepsilon, x)$ function[3]; and the eigenvalues are then determined from a recent series expansion[4] in terms of the Hermite functions for the solution of the differential equation whose exact solution is the aforesaid $HeunB$ function.

Radiative corrections are a commonplace in quantum field theory, a standard example being the calculation and experimental assessment of the corrections to $g - 2$ for the muon[5]; the exercise has been path-breaking in that it has provided a stringent test of the standard model of particle physics and a window to the physics beyond. As the learned reader is doubtless aware, radiative corrections due to gravitation are a different ball-game, it being enough for our purposes to mention two examples here: a. a perturbative calculation [6] about a flat background in two dimensional models of quantum gravity using operator regularization [7], a symmetry – preserving procedure that is particularly convenient as it does not alter the dimensionality of the models; b. loop quantum gravity[8] – a mathematically well defined, non-perturbative and background independent quantisation of general relativity with its conventional matter couplings; amongst its most significant results being a derivation of the Bekenstein – Hawking black hole entropy formula.

This report is the outcome – but nevertheless inspired by the above remarks – of a different exercise, namely to determine how an exact quantum mechanical result is altered by the introduction of the Newtonian gravitational potential. One is not calculating a radiative correction here, but given the infirmities associated with gravitational corrections in quantum field theory a calculation of this kind is not without substance after all. Without hesitation, the results obtained herein encourage one to anticipate a different calculation along similar lines soon.

Consider the setup in the figure; with the indicated entries the Hamiltonian is easily written as

$$H = \frac{p_1^2}{2m_1} + \frac{p_2^2}{2m_2} + \frac{1}{2}k(r_1 - a)^2 + \frac{1}{2}\acute{k}(r_2 - r_1 - a)^2 + \frac{1}{2}k(r_2 - 2a)^2 \qquad (1)$$

with $p_1 = m_1\dot{r}_1$ and $p_2 = m_2\dot{r}_2$ being the canonical momenta of the masses $m_1$ and $m_2$ respectively; in terms of the centre of mass and relative coordinates

$$R = \frac{m_1 r_1 + m_2 r_2}{m_1 + m_2}, r = r_2 - r_1$$

one gets on taking $m_1 = m_2 = m, k = \acute{k}$

$$H = \frac{P^2}{2M} + kR^2 - 3akR + \frac{p^2}{2\mu} + l((r-a)^2 + 3a^2) \qquad (2)$$

with $M = m_1 + m_2 = 2m, M\mu = m_1 m_2 = m^2, l = \frac{3}{4}k$. With the labels $P = M\dot{R}, p = \mu\dot{r}$ for the canonical momenta for the centre of mass and reduced mass respectively the Hamiltonian for the latter is

$$H_\mu = \frac{p^2}{2\mu} + l((r-a)^2 + 3a^2)) \qquad (3a)$$

with the corresponding Schrodinger equation

$$\left(-\frac{\hbar^2}{2\mu}\frac{d^2}{dr^2} + l((r-a)^2 + 3a^2) - E\right)\psi = 0 \qquad (3b)$$

or

$$\frac{d^2\psi}{dq^2} + (c - q^2)\psi = 0 \qquad (4)$$

with

$$E = \frac{1}{2}\lambda\hbar\omega\,,\,\omega = \sqrt{\frac{2l}{\mu}}\,,\,r = \sqrt{\frac{\hbar}{\mu\omega}}\,s,\,a = \sqrt{\frac{\hbar}{\mu\omega}}\,b,\,q \equiv s - b = \sqrt{\frac{\mu\omega}{\hbar}}(r_2 - r_1 - a),\,c = \lambda - 3b^2.$$

The centre of mass motion will be ignored in the sequel as the gravitational potential depends on the relative coordinate $r$ only.

A general solution to (4) is given in terms of the parabolic cylinder function $D_\nu(x)$ by

$$\psi(q) = c_1 D_\nu(\sqrt{2}\,q) + c_2 D_\rho(i\sqrt{2}\,q) \qquad (5)$$

with $2\nu = -1 + c,\, 2\rho = -1 - c$ and $c_1, c_2$ being arbitrary constants. Note that the sum $\nu + \rho = -1$; the boundary conditions on $\psi(q)$ will determine the eigenvalues in the calculation below. The conditions $\psi(q = 0) = 0$ and $\psi(q = -2a) = 0$ seem appropriate as they relate to $r_1 = a, r_2 = 2a$ and $r_1 = 2a, r_2 = a$, the former being the equilibrium position and the latter a position at which the two masses are stretched maximally. Eq.(5) then leads to

$$c_1 D_\nu(0) + c_2 D_\rho(0) = 0\,,\,c_1 D_\nu(z) + c_2 D_\rho(iz) = 0\,,\,z \equiv -2a\sqrt{\frac{2\mu\omega}{\hbar}} \qquad (6)$$

and a nontrivial solution for the $c_1, c_2$ follows when

$$D_\nu(0)D_\rho(iz) - D_\rho(0)D_\nu(z) = 0 \qquad (7)$$

Or

$$\frac{D_\rho(iz)}{D_\rho(0)} = \frac{D_\nu(z)}{D_\nu(0)} \qquad (8)$$

With the recurrence relation[9]

$$D_\rho(iz) = \frac{\Gamma(-\nu)}{i\sqrt{2\pi}\,e^{-i\pi\nu/2}}\left(e^{-i\pi\nu}D_\nu(-z) - D_\nu(z)\right),\,\rho = -\nu - 1 \qquad (9)$$

one gets

$$\left(\frac{D_\rho(0)}{D_\nu(0)} + \frac{\Gamma(-\nu)e^{i\pi\nu/2}}{i\sqrt{2\pi}}\right)D_\nu(z) = \frac{\Gamma(-\nu)e^{-i\pi\nu/2}}{i\sqrt{2\pi}}D_\nu(-z) \qquad (10)$$

and the properties of the gamma function[9] lead to

$$\frac{D_\rho(0)}{D_\nu(0)} = 2^{\rho+\frac{1}{2}}\frac{\Gamma\left(\frac{1-\nu}{2}\right)}{\Gamma\left(\frac{1-\rho}{2}\right)} = 2^{\rho+\frac{1}{2}}\frac{\Gamma\left(\frac{1}{2}+u\right)}{\Gamma(1-u)} = \sqrt{\frac{2}{\pi}}\Gamma(1+\rho)\cos\frac{\pi\rho}{2},\,u = \frac{1+\rho}{2} \qquad (11)$$

so that the bracket on the left hand side of (10) writes as

$$\frac{1}{\sqrt{2\pi}}\Gamma(1+\rho)\left(2\cos\frac{\pi\rho}{2} - ie^{-\frac{i\pi}{2}(\rho+1)}\right) = \frac{1}{\sqrt{2\pi}}\Gamma(1+\rho)e^{i\pi\rho/2} \qquad (12)$$

Eq.(10) thus becomes

$$\frac{1}{\sqrt{2\pi}}\Gamma(1+\rho)e^{i\pi\rho/2}\big(D_\nu(z) - D_\nu(-z)\big) = 0$$

or $\qquad\qquad\qquad D_\nu(z) = D_\nu(-z) \qquad (13)$

Eq.(13) is important because the eigenvalues of the Hamiltonian in (3a) follow easily from either of [9]

$$D_\nu(-z) = D_\nu(z) - 2^{\frac{\nu+1}{2}} z e^{-\frac{z^2}{4}} \Gamma\left(\frac{\nu+1}{2}\right) \sin\frac{\pi\nu}{2} \, L^{\frac{1}{2}}_{\frac{\nu-1}{2}}\left(\frac{z^2}{2}\right) \tag{14a}$$

with $L_\nu^\lambda$ being the generalized Laguerre function, or

$$D_\nu(-z) = D_\nu(z) - 2^{\frac{\nu+1}{2}} \frac{\nu \Gamma\left(\frac{\nu}{2}\right) \sin\frac{\pi\nu}{2}}{\sqrt{\pi}} z e^{-\frac{z^2}{4}} M\left(\frac{1-\nu}{2}; \frac{3}{2}; \frac{z^2}{2}\right) \tag{14b}$$

with $M(a; b; z)$ being the Kummer confluent hypergeometric function. Since $z$ is arbitrary eqs.(13) and (14a) – or (14b) – lead to the eigenvalues as the roots of the equation

$$\sin\frac{\pi\nu}{2} = 0 \tag{15}$$

The definitions $\nu = \frac{-1+c}{2}$, $c = \lambda - 3b^2$ then yield

$$\lambda = 1 + 3b^2 + 4m \tag{16}$$

as the eigenvalues of the Hamiltonian (3a) with $m = 0, \pm 1, \pm 2, \pm 3 \ldots$

Eq.(16) has been derived here without recourse to approximation and is therefore an exact answer, just as the solution to (4) in (5) was exact.

The inclusion of the Newtonian gravitational potential alters the Hamiltonian in (1) by $-G\frac{m_1 m_2}{r}$, eq.(3b) to

$$\left(-\frac{\hbar^2}{2\mu}\frac{d^2}{dr^2} + l((r-a)^2 + 3a^2) - G\frac{m^2}{r} - E\right)\psi = 0 \tag{17}$$

and (4) to

$$\frac{d^2\psi}{ds^2} + \left(c + \frac{K}{s} - (s-b)^2\right)\psi = 0, K \equiv \frac{Gm^3}{\hbar^2}\sqrt{\frac{\hbar}{\mu\omega}} \tag{18}$$

Note that a change of variable from $s$ to $q$ has not been effected in (18) for convenience. To our knowledge eq.(18) cannot be solved exactly; one has to perforce resort to other methods to arrive at answers and in this effort a sequence of transformations seems promising[10]. We begin with $\psi = e^{-\frac{s^2}{2}}\chi$ to get

$$s\frac{d^2\chi}{ds^2} - 2s^2\frac{d\chi}{ds} + (K + (c - 1 - b^2)s + 2bs^2)\chi = 0 \tag{19}$$

and with $\chi = e^{bs}\varphi$ one gets

$$s\frac{d^2\varphi}{ds^2} + 2s(b-s)\frac{d\varphi}{ds} + (K + (c-1)s)\varphi = 0 \tag{20}$$

The definition $c = \lambda - 3b^2$ in (20) does not augur well for a relation between λ and the constant K – the latter is defined in terms of the gravitational constant G as seen from eq.(18) above – from a Frobenius series expansion for $\varphi(s)$ in (20); it is easy to check this explicitly. Perhaps such a relation can be got from the imposition of the boundary conditions on the exact solution to (20).

Eq.(20) bears comparison with

$$x\frac{d^2y}{dx^2} + (\gamma + \delta x + \varepsilon x^2)\frac{dy}{dx} + (\alpha x - q)y = 0 \tag{21}$$

whose solution is the bi-confluent HeunB$(q, \alpha, \gamma, \delta, \varepsilon, x)$ function; as a follow-up one also has

$$\gamma = 0, \delta = 2b, \varepsilon = -2, q = -K \text{ and } \alpha = c - 1 \tag{22}$$

Parenthetically, *Mathematica* yields the general solution to (20) as

$$\varphi(s) = b_1 HeunB(-K, c-1, 0, 2b, -2, s) + b_2 s HeunB(-K-2b, c-3, 2, 2b, -2, s) \tag{23}$$

with $b_1, b_2$ as constants. The general solution to (18) is thus given by

$$\psi(s) = e^{bs - \frac{s^2}{2}}\varphi(s) \tag{24}$$

and the boundary conditions $\psi(s = -a) = 0$, $\psi(s = a) = 0$ lead to

$$e^{-ab - \frac{a^2}{2}}\big(b_1 HeunB(1, -a) - ab_2 HeunB(2, -a)\big) = 0$$

$$e^{ab - \frac{a^2}{2}}\big(b_1 HeunB(1, a) + ab_2 HeunB(2, a)\big) = 0 \tag{25}$$

with the first five arguments of the two terms in (23) being collectively labelled as 1 and 2 respectively. A nontrivial solution to $b_1, b_2$ in (25) is got when

$$HeunB(1, -a)HeunB(2, a) + HeunB(1, a)HeunB(2, -a) = 0$$

or $\qquad\qquad\dfrac{HeunB(1,-a)}{HeunB(1,a)} = -\dfrac{HeunB(2,-a)}{HeunB(2,a)} \tag{26}$

thus making eq.(26) the counterpart of (8) above. The literature on the HeunB function [Hortacsu[3],Ref.9] is of little help in eliciting the desired answer from (26) and one is left to pry the result on one's own; we therefore ignore eq.(26) and use the relation [9]

$$H_\nu(z) = 2^{\frac{\nu}{2}} e^{\frac{z^2}{2}} D_\nu(\sqrt{2}z) \tag{27}$$

between the Hermite functions $H_\nu(z)$ and the parabolic cylinder functions $D_\nu(z)$ in (27) in an expansion of the solution of eq.(21) in terms of the Hermite functions for the answer, it being inspired by Ishkhanyan and Ishkhanyan[4].To elaborate, as $H_\nu(z)$ are solutions[9] to

$$\frac{d^2y}{dz^2} - 2z\frac{dy}{dz} + 2\nu y = 0 \tag{28}$$

on writing

$$y(x) = \sum_0^\infty c_n u_n, u_n = H_{\alpha_n}(t_0(x + x_0)), \alpha_n = \alpha_0 + n \tag{29}$$

where $\alpha_0, t_0$ and $x_0$ are constants to be fixed below, one gets the Hermite functions in (29) as solutions to

$$\frac{d^2 u_n}{dx^2} - 2t_0^2(x + x_0)\frac{du_n}{dx} + 2t_0^2 \alpha_n u_n = 0 \qquad (30)$$

and the use of (29) and (30) in (21) yields

$$\sum_0^\infty c_n \left[ ((2t_0^2 + \varepsilon)x^2 + (\delta + 2t_0^2 x_0)x + \gamma)\frac{du_n}{dx} + (\alpha x - q - 2xt_0^2 \alpha_n)u_n \right] = 0 \qquad (31)$$

With (22) and the choice $2t_0^2 = -\varepsilon = 2$, the coefficient of the first derivative term in (31) becomes $(\delta + 2x_0)x$ and one can now set $x_0 = 0$ without loss of generality as $\delta \neq 0$. As a bonus the argument of the Hermite function in (29) is now $t_0 x$ making eq.(27) tantalizingly useful with $t_0^2 = 1$. The use of the recurrence identities [9]

$$\frac{du_n}{dx} = 2t_0 \alpha_n u_{n-1}, \quad t_0 x u_n = \alpha_n u_{n-1} + \frac{1}{2} u_{n+1} \qquad (32)$$

rewrites (31) first as

$$\sum_0^\infty c_n [2\delta \alpha_n t_0 x u_{n-1} + (\alpha - 2\alpha_n)x u_n - q u_n] = 0 \qquad (33)$$

and then as

$$\sum_0^\infty c_n \left[ 2\delta t_0 \alpha_n \alpha_{n-1} u_{n-2} + t_0(\delta \alpha_n - q)u_n + (\alpha - 2\alpha_n)\left(\alpha_n u_{n-1} + \frac{1}{2} u_{n+1}\right) \right] = 0 \qquad (34)$$

Eq.(34) yields a 4-term recurrence relation given by

$$2\delta t_0 \alpha_n \alpha_{n-1} c_n + (\alpha - 2\alpha_{n-1})\alpha_{n-1} c_{n-1} + t_0(\delta \alpha_{n-2} - q)c_{n-2} + \frac{1}{2}(\alpha - 2\alpha_{n-3})c_{n-3} = 0 \qquad (35)$$

and is the counterpart of eq.(8) in Ref.4. Since $\alpha_n = \alpha_0 + n$ and $H_{\alpha_n}$ are polynomials of degree $\alpha_n$ when $\alpha_n$ are integers one should look for noninteger values for the last parameter $\alpha_0$ that remains to be fixed among the three in (29), given that $x_0, t_0$ have been dealt with above; it would thus be naïve to take $c_n = 0$ for $n = -1, -2,$ and $-3$ in (35). Instead the zeros of $\alpha - 2\alpha_0$ and $\delta \alpha_0 - q$ seem plausible with the latter a better choice given that the order of $H_{\alpha_n}$ and hence – the sought after eigenvalues – would then depend on the gravitational constant $K$ as $q + K = 0$. The expansion in (29) would then become with $\delta = 2b$

$$y(x) = \sum_0^\infty c_n u_n, \quad u_n = H_{\alpha_n}(t_0 x), \quad \alpha_n = \alpha_0 + n, \quad \alpha_0 = -\frac{K}{2b} \qquad (36)$$

The contrary choice $\alpha_0 = \frac{\alpha}{2}$ would lead us back to eq.(16) as $\alpha = c - 1$ vide eq.(22) with little to show for the effort invested above thus making the value of $\alpha_0$ in (36) unique by default; additionally one now has from (35) $c_n = 0$ for $n = -1, 1$ and $2$ thus reworking (36) to

$$y(x) = c_0 u_0 + \sum_3^\infty c_n u_n, \quad u_n = H_{\alpha_n}(t_0 x), \quad \alpha_n = \alpha_0 + n, \quad \alpha_0 = -\frac{K}{2b} \qquad (37)$$

with $c_0$ arbitrary and $c_3 = -c_0 \frac{\alpha - 2\alpha_0}{4\delta t_0 \alpha_3 \alpha_2}$. The $c_n, n > 3$ can also be easily got from (35) with the necessary inputs; for instance,

$$c_4 = -c_3 \frac{(\alpha - 2\alpha_3)}{2\delta t_0 \alpha_4}, \quad c_5 = -c_4 \frac{(\alpha - 2\alpha_4)}{2\delta t_0 \alpha_5} - c_3 \frac{(\delta \alpha_3 - 2q)}{2\delta \alpha_4 \alpha_5} \qquad (38)$$

To tie up the aforesaid discussion with eq.(20) we return to eq.(24) and write

$$\psi(s) = e^{bs - \frac{s^2}{2}} \varphi(s), \varphi(s) = c_0 u_0 + \sum_3^\infty c_n u_n, u_n = H_{\alpha_n}(t_0 s), \alpha_n = \alpha_0 + n, \alpha_0 = -\frac{K}{2b} \quad (39)$$

or,

$$\psi(s) = e^{bs - \frac{s^2}{2}} \sum_0^\infty c_n H_{\alpha_n}(t_0 s) \quad (40)$$

it being understood that $c_1, c_2$ are zero. The conditions $\psi(s = -a) = 0$, $\psi(s = a) = 0$ thus appear as

$$\sum_0^\infty c_n H_{\alpha_n}(-t_0 a) = 0, \quad \sum_0^\infty c_n H_{\alpha_n}(t_0 a) = 0 \quad (41)$$

or with (27) as

$$\sum_0^\infty c_n D_{\alpha_n}(-\sqrt{2} t_0 a) = 0, \quad \sum_0^\infty c_n D_{\alpha_n}(\sqrt{2} t_0 a) = 0 \quad (42)$$

The first of eqs.(42) can be reworked using eq.(14a) as

$$\sum_0^\infty c_n D_{\alpha_n}(-w) = \sum_0^\infty c_n D_{\alpha_n}(w) - w e^{-\frac{t_0^2 a^2}{2}} \sum_0^\infty c_n 2^{\frac{\nu+1}{2}} \Gamma\left(\frac{\nu+1}{2}\right) \sin\frac{\pi\nu}{2} L_{\frac{\nu-1}{2}}^{\frac{1}{2}}(t_0^2 a^2) \quad (43)$$

with $w = \sqrt{2} t_0 a, \nu = \alpha_n$. With the left hand side and the first sum on the right respectively equal to zero by eq.(42), (43) becomes

$$\sum_0^\infty c_n 2^{\frac{\nu+1}{2}} \Gamma\left(\frac{\nu+1}{2}\right) \sin\frac{\pi\nu}{2} L_{\frac{\nu-1}{2}}^{\frac{1}{2}}(a^2) = 0 \quad (44)$$

recalling that $t_0^2 = 1$. Eq. (44) is an infinite series containing the constants $c_n$ which are zero for $n = 1,2$ and nonzero otherwise; given that $a$ is a constant the location of the zeros of the $L_{\frac{\nu-1}{2}}^{\frac{1}{2}}(a^2)$ as a function of $\nu$ could not be accessed in the literature[9] easily, thus leaving $\sin\frac{\pi\nu}{2} = 0$ as the only option, that is

$$\nu = 2m, m = 0, \pm 1, \pm 2 \ldots \text{ or } \alpha_n = \alpha_0 + n = 2m \quad (45)$$

or

$$\alpha_0 = -\frac{K}{2b} = 2m - n$$

Since $K$ and $2b$ are each nonnegative (45) reworks to

$$\frac{K}{2b} = |n - 2m|, n = 0,1,2 \ldots, m = 0, \pm 1, \pm 2 \ldots \quad (46)$$

Eq.(46) gives the 'eigenvalues' associated with eq.(20) and is the counterpart of eq.(16) with the eigenfunctions given by (23).

In conclusion, the rarefied nature of the HeunB functions has pre-empted a derivation of the eigenvalues beginning with (26) here; that is a task for the future.

A mention should be made here that Alberg and Wilets[11] have reported the determination of exact solutions to the Schrodinger equation for potentials containing the Coulomb($\sim 1/r$) plus harmonic oscillator($\sim r^2$) term but subject to a constraint on the ratio between the strengths of the Coulomb and harmonic oscillator terms. The calculations in this paper are exact and not impaired by this latter limitation; also our workout is in 1-space dimension unlike Ref.11 where the reference is to spherically symmetric potentials with the centrifugal term $\frac{l(l+1)}{r^2}$.

The author has no conflicts to disclose.

**Acknowledgement:** The author thanks the referee for encouraging comments.

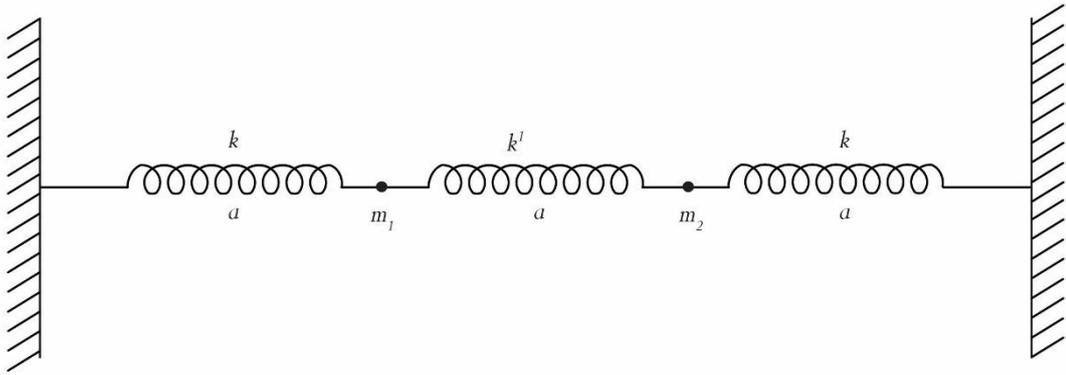

Fig. The masses $m_1$ and $m_2$ have instantaneous positions $r_1$ and $r_2$ respectively as measured from the left wall along the line joining the two masses with $r_2 > r_1$; at equilibrium $r_1 = a$ and $r_2 = 2a$, $a$ being the length of each spring in the unstretched configuration. The force constants are also shown.